\documentclass[12pt]{iopart} 
\usepackage{iopams}   
\usepackage{graphicx,cite,epsf} 
\usepackage{psfrag} 
\begin{document} 

\title[Rosette versus star-like structures]{Conformational properties of complex
polymers: rosette versus star-like structures} 

\author{V. Blavatska} 
\address{Institute for Condensed Matter Physics of the National Academy of Sciences of Ukraine, 
79011 Lviv, Ukraine} 
\ead{viktoria@icmp.lviv.ua} 
\author{R. Metzler} 
\address{Institute for Physics and Astronomy, University of Potsdam, 14476 Potsdam-Golm, Germany} 
\address{Department of Physics, Tampere University of Technology, FI-33101 Tampere, Finland}

\begin{abstract} 
Multiple loop formation in polymer macromolecules is an important feature of the chromatin 
organisation and DNA compactification in the nuclei.   
We analyse the size and shape characteristics of complex polymer structures,   
containing in general $f_1$ loops (petals) and $f_2$ linear chains (branches). 
Within the frames of continuous model of Gaussian macromolecule, we apply the path integration method 
and obtain the estimates for gyration radius $R_g$ and asphericity $\hat{A}$ of typical conformation 
as functions of parameters $f_1$, $f_2$. In particular, our results qualitatively reveal the extent of anisotropy 
of star-like topologies as compared to the rosette structures of the same total molecular weight. 
\end{abstract} 

%{\it Keywords}: polymers, path integration, conformational properties 
\submitto{\JPA} 
\section{Introduction} 

Loop formation in macromolecules plays an important role in a number of biochemical processes: stabilisation of globular proteins  \cite{Perry84,Wells86,Pace88,Nagi97}, 
transcriptional regularisation of genes \cite{Schlief88,Rippe95,Towles09} as well as    
DNA compactification in the nucleus \cite{Fraser06,Simonis06,Dorier09}. 
The localisation of chromatin fibres to semi-compact regions known as 
chromosome territories  is maintained among others by 
the topological constraints introduced by multiple loops in chromatin organisation \cite{Nooijer09}. 
Numerous analytical and numerical studies have been conducted to analyse the cyclisation probability and 
loop size distributions in long flexible macromolecules \cite{Chan89,Rey91,Rubio93,Wittkop96,Zhou01,Cui06,Toan06,Shin14}. 
The conformational properties of isolated loops (ring polymers) \cite{Bishop86,Diel89,Jagodzinski92,Obukhov94,Muller99,Deutsch99,Miyuki01,Calabrese01,Alim07,Bohr10,Sakaue11,Jung12,Ross11} and 
multiple loops \cite{Dorier09,Shin14a} have been intensively studied as well. 

\begin{figure}[t!] 
\begin{center} 
\includegraphics[width=12cm]{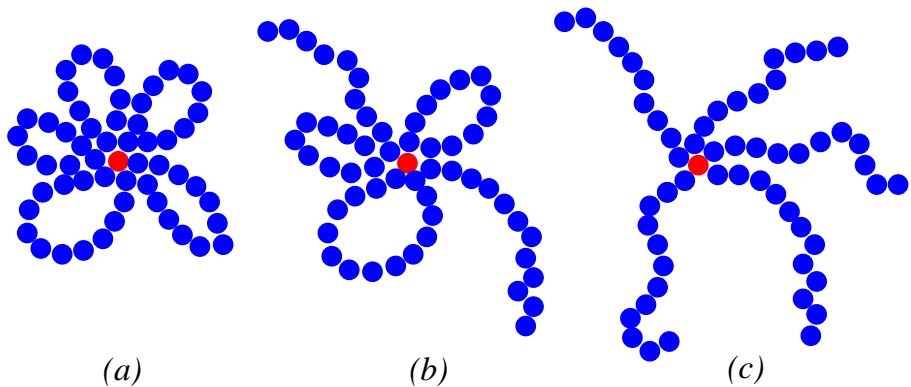}
\end{center} 
\caption{Schematic presentation of polymer systems of complex topologies.} 
\label{structura} 
\end{figure} 

Interlocking and entanglements are ubiquitous features of flexible polymers of high molecular weight. 
In particular,  DNA  can exist in the form of catenated (bonded) rings of various complexities \cite{Cluzel96,Yamaguchi00,Lyubchenko02}.
The segments of different DNA molecules can intercross through the transient breaks introduced by  special enzymes (topoisomerases)  \cite{Wang85,Champoux,Krasnow81}. 
%The study of proteins consisting of interlocked circular chains both with links 
%formed by covalently bonded protein subunits (bonded or knotted rings) 
%\cite{Duda98,Zhou03,Metzler02,Metzler02a,Metzler07} and topological 
%links (catenated rings) \cite{Zhou03,Wikoff00,Blankenship03,Bohn10} 
Numerous studies \cite{Duda98,Zhou03,Metzler02,Metzler02a,Metzler07,Wikoff00,Blankenship03,Bohn10}
reveal 
the advantage of linking in stabilisation of peptide and 
protein oligomers. 

In this concern, it is worthwhile to study the conformational properties of generalised complex polymer structures, 
containing $f_1$ loops and $f_2$ linear chains of the same length $S$(see Fig. \ref{structura}). 
The  properties of so-called ``rosette'' (Fig. \ref{structura}a) and ``ring-linear'' (Fig. \ref{structura}b) structures 
have been considered recently in numerical simulations in Refs. \cite{Nooijer09,Bohn10a}. 
{However, only the shape properties of very simple structures of
two bonded rings ($f_2=2, f_1=0$) and rings with two connected linear branches
($f_2=1, f_1=2$) with excluded volume effects were analysed \cite{Bohn10a}. In
particular, a quite subtle difference in the conformational properties of two
connected polymer rings compared with that of one isolated ring were found.}
On the other hand, the properties of ``star'' polymers (Fig. \ref{structura}c)
have been intensively studied by now \cite{Duplantier89,Shafer92,Grest87,Hsu04,Ohno91,Shida96,Ferber95,Ferber96,Ferber02}.   

In general, the statistics of polymers is known to be characterised by a set of universal properties, which are
independent of the  details of the microscopic chemical structure \cite{deGennes,desCloiseaux}. 
In particular, 
the asymptotic number of possible conformations of  structure shown in Fig. \ref{structura}b  obeys the scaling law \cite{Duplantier89} 
\begin{equation} 
{\cal Z}_{f_1f_2}\sim \mu^{(f_1+f_2)S}S^{\gamma_{f_1f_2}-1},\,\,\, 
\gamma_{f_1f_2}=1-d\nu  f_1+\sigma_{2f_1+f_2}+f_2\sigma_{1}. \label{scalingZ} 
\end{equation} 
Here, $\nu$ is the universal  critical (Flory) exponent, governing the scaling of the size measure (e.g. the averaged gyration radius $R_{g}$) 
of  macromolecule according to
\begin{equation} 
\langle R_{g\,f_1f_2}^2 \rangle \sim  ((f_1+f_2)S)^{2\nu}, 
\end{equation}   
$\sigma_{i}$ is the set of so-called vertex exponents (with $\sigma_2=0$),   
$d$ is the spatial  dimension and $\mu$ is a non-universal fugacity. For an idealised Gaussian (phantom) macromolecule 
without any interactions between monomers one finds: $\nu=1/2$, $\sigma_i=0$. In the limiting cases ($f_1=0$) and ($f_2=0$), respectively, we 
obtain the corresponding critical exponents, governing the scaling of the number of possible conformations of 
 $f_2$-branch star and $f_1$-petal rosette polymers: $\gamma_{{\rm star}}=1$, $\gamma_{{\rm rosette}}=1-df_1/2$. 
Therefore, additional topological constraints 
in rosette polymers lead to a considerable reduction of the number of allowed conformations, 
 as compared with star polymers of the same molecular weight.   

To compare the size measures of macromolecules of different topologies but of the same total molecular weight 
one can consider the  universal  size ratios. 
In particular, in the idealised Gaussian case, the ratio of the gyration radii of the individual ring and open linear structures reads  \cite{Zimm49} 
\begin{eqnarray} 
&&\frac{\langle R_{g\,{\rm ring}}^2\rangle}{\langle R_{g\,{\rm linear}}^2\rangle}=\frac{1}{2},\label{ratioRL} 
\end{eqnarray} 
whereas comparing the size of $f$-branch star and a linear chain of the same total length one has \cite{Zimm49} 
\begin{eqnarray} 
 &&\frac{\langle R_{g\,f{\rm star}}^2\rangle}{\langle R_{g\,{\rm linear}}^2\rangle}=\frac{3f-2}{f^2}. \label{ratiostar} 
\end{eqnarray} 
Taking into account the excluded volume effect leads to an increase of this values \cite{Baumgaertner81,Prentis82,Miyake82,Alessandrini92,Grest87,Bishop93,Wei97}. 

The overall shape of a typical polymer conformation is of great importance,  affecting in particular the mobility and 
folding dynamics of proteins \cite{Dima04,Rawat09}. The shape of DNA may be relevant for the accessibility for enzymes depending on the 
spatial distance between DNA-segments \cite {Hu06}. Already in 1934 it was realised \cite{Kuhn34} that 
the shape of a typical flexible polymer coil in a solvent is  anisotropic and resembles that of a prolate ellipsoid. 
It is convenient to characterise the asymmetry of polymer configurations in terms 
of rotationally invariant universal quantities 
\cite{Aronovitz86,Rudnick86} constructed as combinations of the components of the gyration tensor,  such as the asphericity $ \hat{A} $. 
This quantity takes on a maximum value of unity for a completely stretched, rod-like configuration, 
and equals zero for the spherical form, thus obeying the inequality 
$ 0\leq \hat{A} \leq 1$. 
In the Gaussian case, for the individual linear and circular polymer chains one has correspondingly \cite{Rudnick86} 
\begin{eqnarray} 
&&\hat{A}_{{\rm linear}}=\frac{2(d+2)}{5d+4}, \label{Adchain}\\ 
&&\hat{A}_{{\rm ring}}=\frac{d+2}{5d+2}.\label{Adring} 
\end{eqnarray} 
whereas the asphericity of $f$-branch star polymer 
reads \cite{Wei90}: 
\begin{equation} 
\hat{A}_{{\rm star}} = 2\frac{(2+d)(15f-14)}{5d(3f-2)^2+4(15f-14)}. \label{Adstar} 
\end{equation} 
Note that Eq. (\ref{Adchain}) gives the asphericity of a trajectory of diffusive randomly walking particle.
The influence of excluded volume effects on the shape parameters of single linear and ring polymers, as well as star polymers have been 
analysed so far both analytically 
\cite{Benhamou85,Aronovitz86,Jagodzinski92} and  numerically \cite{Domb69,Solc71,Bishop88,Honeycutt88,Drube10}. 

In the present paper, we study the universal size and shape characteristics of  complex polymer structures (Fig. \ref{structura}), applying the path integration method. A special attention is paid to the analytical study of statistical properties of rosette (Fig. \ref{structura}a) 
and ring-linear (Fig. \ref{structura}b) structures. 

The layout of the paper is as follows. In section 2, we shortly describe the presentation of complex polymer system within the frames of continuous chain model.    Our results are given in Section 3. We end up by giving conclusions in Section 4. 

\section{The model} 

We consider a system consisting of $f_1$ closed polymer loops (petals) and $f_2$ linear chains (branches), all bonded together  at one 
``branching" point (see Fig. \ref{structura}b). 
Within the Edwards continuous chain model \cite{Edwards}, each of the individual branches or petals is presented 
as a path of length $S_i$, parameterised by $\vec{r}_i(s)$, where $s$ is varying from $0$ to $S_i$ ($i=1,2,\ldots,f_1+f_2$). 
For simplicity we take: $S_1=\ldots=S_{f_1+f_2}=S$. The weight of the individual $i$th path is given by
\begin{eqnarray} 
W_i=\exp \left(-\frac{1}{2} \int^{S}_{0}\left(\frac{d\vec{r}_i(s)}{ds}\right)^{2}\!ds \right)
\end{eqnarray} 
and the partition function of the system  can thus be written as 
\begin{eqnarray} 
{\cal Z}_{f_1f_2}=\frac{ {\displaystyle{\int}}\! {\cal {D}}\vec{r}\,\, {\displaystyle{\prod\limits_{j=1}^{f_1}}}\delta(\vec{r}_j(S)-\vec{r}_j(0)) {\displaystyle {\prod\limits_{i=1}^{f_1+f_2}}}\delta(\vec{r}_i(0)) 
W_i} 
{{\displaystyle {\int}}\! {\cal {D}}\vec{r}\,\,{\displaystyle{\prod\limits_{i=1}^{f_1+f_2}}}\delta(\vec{r}_i(0)) 
W_i}. \label{model-con} 
\end{eqnarray} 
Here, $\int\! {\cal D}{\vec r}$ denotes functional path integrations over $f_1+f_2$ trajectories. The products of $\delta$-functions describe the fact that $f_1$ trajectories are closed and that 
the starting point of all trajectories is fixed (the ``branching" point).   
Note, that (\ref{model-con}) is  normalised in such a way that 
the partition function of the system consisting of $f_1+f_2$ open linear Gaussian chains (star-like structure) is unity. 
  
  Exploiting the Fourier-transform of the  $\delta$-functions
\begin{equation} 
\delta (\vec{r}_j(S)-\vec{r}_j(0)) =(2\pi)^{-d}\int {\rm d}\vec{q}_j\, {\rm e}^{-i\vec{q}_j(\vec{r}_j(S)-\vec{r}_j(0))} \label{d} 
\end{equation} 
and rewriting in the exponent
\begin{eqnarray} 
{-}\frac{1}{2}\int\limits_{0}^{S}\left(\frac{{\rm d} {\vec r}_{j}(s)}{{\rm d} s}\right)^2\!{\rm d}s 
- 
i\vec{q}_j\!\int\limits_{0}^{S}\frac{d \vec{r}_j(s)}{d\,s}{\rm d}s = 
{-}\frac{1}{2}\int\limits_{0}^{S}{\rm d}s\left( 
\left(\frac{{\rm d} {\vec r}_{j}(s)}{{\rm d} s}+i\vec{q}_j\right)^2 \!{+} 
q^2 \right) \nonumber 
\end{eqnarray} 
we evaluate the expression of partition function (\ref{model-con}), giving the asymptotic number of 
possible conformations of polymer system
\begin{eqnarray} 
&&{\cal Z}_{f_1f_2}= 
%\frac{1}{Z^0} \left[{\cal {D}}\vec{r}\,\,\delta(\vec{r}(0)){\rm e} ^{-\frac{1}{2}\int_0^S ds\left(\frac{d\vec{r}(s)}{ds}\right)^2}\right]^{f_1+f_2}\! 
(2\pi)^{-f_1d}\prod\limits_{j=1}^{f_1}\int {\rm d}\vec{q}_j\, {\rm e}^{-\frac{q_j^2S}{2}}= (2\pi S)^{-df_1/2}. 
\end{eqnarray} 
Comparing this relation with Eq. (1), we  recover the estimate for the critical exponent $\gamma_{f_1f_2}=1-df_1/2$.      

The average of any observable $\langle (\ldots) \rangle$ over an ensemble of conformations is then given by:
\begin{eqnarray} 
\langle (\ldots) \rangle= \frac{1}{{\cal Z}_{f_1f_2}}\frac{ {\displaystyle{\int}}\! {\cal {D}}\vec{r}\,\, {\displaystyle{\prod\limits_{j=1}^{f_1}}}\delta(\vec{r}_j(S)-\vec{r}_j(0)) {\displaystyle {\prod\limits_{i=1}^{f_1+f_2}}}\delta(\vec{r}_i(0)) 
(\ldots) W_i} 
{{\displaystyle {\int}}\! {\cal {D}}\vec{r}\,\,{\displaystyle{\prod\limits_{i=1}^{f_1+f_2}}}\delta(\vec{r}_i(0)) 
W_i}. 
\label{modelav} 
\end{eqnarray}

\section{Size ratios and asphericity} 

The size and shape characteristics of a typical polymer conformation can be characterised 
 \cite{Solc71} in terms of the gyration tensor $\bf{Q}$. Within the framework of a continuous polymer model 
 the  components of this tensor can be presented as 
\begin{equation} 
\hspace*{-1cm}Q_{\alpha\beta}=\frac{1}{2S^2(f_1+f_2)^2} \sum_{i,j=1}^{f_1+f_2}\int_0^S\!\! d s_1 \!\int_0^{S}\!\! d s_2 \,\, 
( {\vec{r}_i}^{\,\alpha}( s_2){-}\vec{r}_j^{\,\alpha}( s_1 )) ( \vec{r}_i^{\,\beta}( s_2){-}\vec{r}_j^{\,\beta}( s_1 ) ), \label{mom} 
\end{equation} 
where ${r}_i^{\alpha}( s_1)$ is $\alpha$th  component of $\vec{r}_i( s_1)$ ($\alpha=1,2,\ldots,d$). 

For the  averaged radius of gyration one has 
\begin{eqnarray} 
&&\langle R_{g\,f_1f_2}^2 \rangle = \frac{1}{2S^2(f_1+f_2)^2} \sum_{i,j=1}^{f_1+f_2} \left\langle  \int_0^S \!\! d s_1\int_0^{ s_1} \!\! d s_2 
(\vec{r}_i( s_2)-\vec{r}_j( s_1 ))^2 \right\rangle \label{rg} \\ 
&&= \langle {\rm Tr}\,\bf{Q} \rangle. \nonumber 
\end{eqnarray} 
Here and below, $\langle(\ldots)\rangle$ denotes averaging over an ensemble of path conformations according to (\ref{modelav}).

The spread in the eigenvalues $\lambda_i$ of the gyration tensor (\ref{mom}) describes the distribution of monomers inside the polymer coil and 
thus measures the asymmetry of the molecule. For a symmetric (spherical) 
configuration all the eigenvalues $\lambda_{i}$ are equal, whereas for completely stretched rod-like conformation all $\lambda_{i}$ are zero except one. 
Let ${\overline{\lambda}}\equiv {\rm Tr}\, {\bf{Q}}/d$ 
be the mean eigenvalue of  the gyration tensor. 
Then one may characterise the extent of asphericity of a macromolecule by the quantity ${{\hat{A}}}$ defined as \cite{Aronovitz86}: 
\begin{equation} 
{\hat{A}} =\frac{1}{d(d{-}1)} \sum_{i=1}^d\frac{\langle(\lambda_{i}{-}{\overline{\lambda}})^2\rangle}{\langle\overline{\lambda}^2\rangle}= 
\frac{d}{d{-}1}\frac{\langle \rm {Tr}\,\bf{\hat{Q}}^2\rangle}{\langle(\rm{Tr}\,{\bf{Q}})^2\rangle}, \label{add} 
\end{equation} 
with ${\bf{{\hat{Q}}}}\equiv{\bf{Q}}{-}\overline{\lambda}\,{\bf{I}}$ (here $\bf{I}$ is  the unity matrix). 
This universal quantity equals zero when $\lambda_i=\overline{\lambda}$, and takes a maximum value 
of one in the case of a rod{-}like configuration. 
The asphericity (\ref{add}) can be rewritten in terms of the averaged components of gyration tensor (\ref{mom}) as follows 
\cite{Aronovitz86}: 
\begin{equation} 
\hat{A}=\frac{\langle Q_{\alpha\alpha}Q_{\alpha\alpha} \rangle+d \langle Q_{\alpha\beta}Q_{\alpha\beta}\rangle -\langle Q_{\alpha\alpha}Q_{\beta\beta} \rangle}{\langle Q_{\alpha\alpha}^2 \rangle+ 
(d-1)\langle Q_{\alpha\alpha}Q_{\beta\beta} \rangle}.\label{A} 
\end{equation} 

Below, we give detailed evaluation of the expressions for the averaged radius of gyration  (\ref{rg}) and  
asphericity (\ref{A}) of the model (\ref{model-con}) within the framework of path integration approach.

\subsection{Radius of gyration} 

\begin{figure}[t!] 
\begin{center} 
\includegraphics[width=13cm]{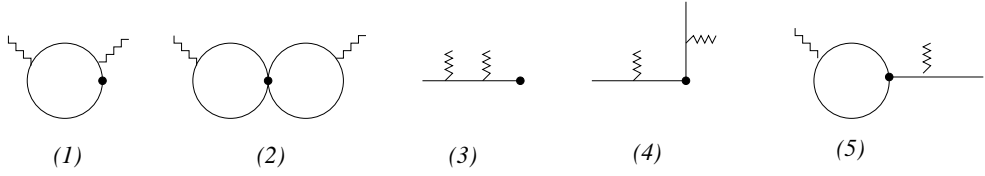} 
\end{center} 
\caption{Diagrammatic presentation of contributions into $\langle \xi(\vec{k}) \rangle$. The solid line on a diagram is a schematic presentation of a polymer path of  length $S$, and 
wavy lines denote so-called restriction points $ s_1$ and $ s_2 $. 
With $\bullet$ we denote the position of the starting ``branching'' point.} 
\label{rgdiagram} 
\end{figure} 

The radius of gyration  (\ref{rg})  can be calculated 
from the identity 
\begin{equation} 
(\vec{r}_i( s_2)-\vec{r}_j( s_1 ))^2=-2{\rm d}\frac{{\partial}}{{\partial} |k|^2} \xi(\vec{k}) \Big|_{k=0},\,\,\,\xi(\vec{k})\equiv{\rm e}^{i\vec{k}\vec{r}_i( s_2)-\vec{r}_j( s_1 )},\label{def} 
\end{equation} 
 and evaluating $\langle \xi(\vec{k}) \rangle$ in the path integration approach.   
In calculation of the contributions into $\langle \xi(\vec{k}) \rangle$ it is convenient  to use
 the diagrammatic presentation as given in Fig. \ref{rgdiagram}. 
According to 
the general rules of diagram calculations \cite{desCloiseaux}, each segment between any two restriction points $s_a$ and $s_b$ 
is oriented and bears a wave vector $\vec{q}_{ab}$ given by a sum of incoming and outcoming  wave vectors injected 
at restriction points and end points. 
At these points, the flow of wave vectors is 
conserved. 
A factor $\exp\left(-{\vec{q}_{ab}}^{\,\,2}(s_b-s_a)/2\right)$ is associated with each segment. An integration is to be made 
over all independent segment areas and over wave vectors injected at the end points. 

The analytic expressions, corresponding to the diagrams (1)-(5) in  Fig. \ref{rgdiagram} then read
\begin{eqnarray} 
\langle \xi(\vec{k}) \rangle_{(1)}&=& \left(\frac{S}{2\pi}\right)^{\frac{d}{2}}\!{\rm e}^{-\frac{{k}^2}{2}(s_2-s_1)}\int {\rm d}q_1\, {\rm e}^{-\frac{{q_1}^2}{2}S}{\rm e}^{-{kq_1}(s_2-s_1)}=\nonumber\\ 
&=&{\rm e}^{-\frac{{k}^2}{2}\frac{(s_2-s_1)(S-s_2+s_1)}{S}},\\ 
\langle \xi(\vec{k}) \rangle_{(2)}&=&\left(\frac{S}{2\pi}\right)^{d}{\rm e}^{-\frac{{k}^2}{2}(s_2+s_1)}\!\int {\rm d}q_1\int {\rm d}q_2\, {\rm e}^{-\frac{{q_1}^2}{2}S}{\rm e}^{-\frac{{q_2}^2}{2}S}{\rm e}^{-q_1k s_1}{\rm e}^{-q_2k s_2}=\nonumber\\ 
&=&{\rm e}^{-\frac{{k}^2}{2}\frac{S(s_1+s_2)-s_1^2-s_2^2}{S}},\\ 
\langle \xi(\vec{k}) \rangle_{(3)}&=&{\rm e}^{-\frac{{k}^2}{2}(s_2-s_1)},\\ 
\langle \xi(\vec{k}) \rangle_{(4)}&=& {\rm e}^{-\frac{{k}^2}{2}(s_2+s_1)},\\ 
\langle \xi(\vec{k}) \rangle_{(5)}&=& \left(\frac{S}{2\pi}\right)^{\frac{d}{2}}{\rm e}^{-\frac{{k}^2}{2}(s_2+s_1)}\int {\rm d}q_1\, {\rm e}^{-\frac{{q_1}^2}{2}S}{\rm e}^{-kq_1s_1}= 
{\rm e}^{-\frac{{k}^2}{2}\frac{S(s_1+s_2)-s_1^2}{S}}. 
\end{eqnarray} 
Taking the derivatives 
with respect to $k$ according to (\ref{def}) in the expressions  above and taking into account the combinatorial factors, 
we find for the radius of gyration 
\begin{eqnarray} 
&&\langle R_{g\,f_1f_2 }^2\rangle  =\frac{Sd}{(f_1+f_2)^2S^2}\left( f_1 \int_{0}^{S}ds_2\int_{0}^{s_2}ds_1 \frac{(s_2-s_1)(S-s_2+s_1)}{S} + \right.\nonumber\\ &&+\frac{f_1(f_1-1)}{2}\int_{0}^{S}ds_2\int_{0}^{S}ds_1 \frac{(s_2-s_1)(S(s_1+s_2)-s_1^2-s_2^2)}{S}  +\nonumber\\ 
&& +f_2\int_{0}^{S}ds_2\int_{0}^{s_2}ds_1(s_2-s_1)+\frac{f_2(f_2-1)}{2} \int_{0}^{S}ds_2\int_{0}^{S}ds_1(s_2+s_1)+ \nonumber\\ 
&&\left. + f_1f_2 \int_{0}^{S}ds_2\int_{0}^{S}\frac{(s_2-s_1)(S(s1+s_2)-s_1^2)}{S}\right)=\nonumber\\ 
&&= \frac{Sd}{(f_1+f_2)^2}\frac{1}{12}\left[ f_1(2f_1-1)+2f_2(3f_2-2)+8f_1f_2\right]. \label{R12} 
\end{eqnarray} 
The case $f_2=0$ corresponds to the rosette structure (Fig. \ref{structura}a) with 
\begin{eqnarray} 
\langle R_{g\, {\rm rosette}}^2 \rangle = \frac{Sd}{12f_1}(2f_1-1),          
\end{eqnarray}   
at $f_1=1$ one recovers  the  gyration radius of individual ring polymer. 
The case $f_1=0$ corresponds to star structure (Fig. \ref{structura}c) with 
\begin{eqnarray} 
\langle R_{g\, {\rm star}}^2 \rangle= \frac{Sd}{6f_2}(3f_2-2), 
\end{eqnarray}   
at $f_2=1$ one receives the gyration radius of linear polymer chain. 

\begin{figure}[b!] 
\begin{center} 
\includegraphics[width=7.5cm]{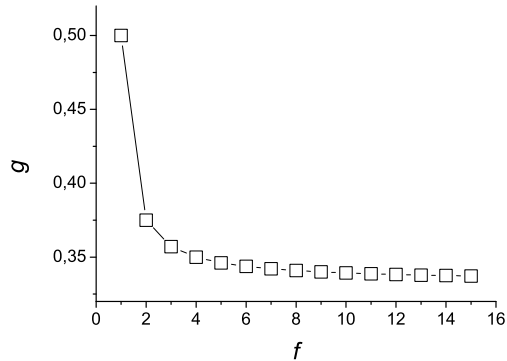} 
\end{center} 
\hspace*{0.5cm} 
\caption{ \label{RRall}  Size ratio (\ref{ratioRS}) of rosette and star polymers of the same total molecular weight as function of branching parameter $f$. Line is guide to the eyes.} 
\end{figure} 

For the size ratio of rosette and star polymers of the same total molecular weight (corresponding to  
$f_1=f_2=f$) we obtain 
\begin{equation} 
g\equiv\frac{\langle R_{g\, {\rm rosette}}^2 \rangle}{\langle R_{g\, {\rm star}}^2 \rangle}=\frac{1}{2}\frac{2f-1}{3f-2}. 
\label{ratioRS} 
\end{equation} 
Note that putting $f=1$ in above relation, one restores the size ratio of the ring and open linear chain of the same molecular weight (\ref{ratioRL}). 
The quantity (\ref{ratioRS}) decreases with increasing the branching parameter $f$ and in the limit   
$f\to \infty$ reaches the asymptotic value $1/3$ (see Fig. \ref{RRall}).

\subsection{Asphericity} 

\begin{figure}[t!] 
\begin{center} 
\includegraphics[width=13cm]{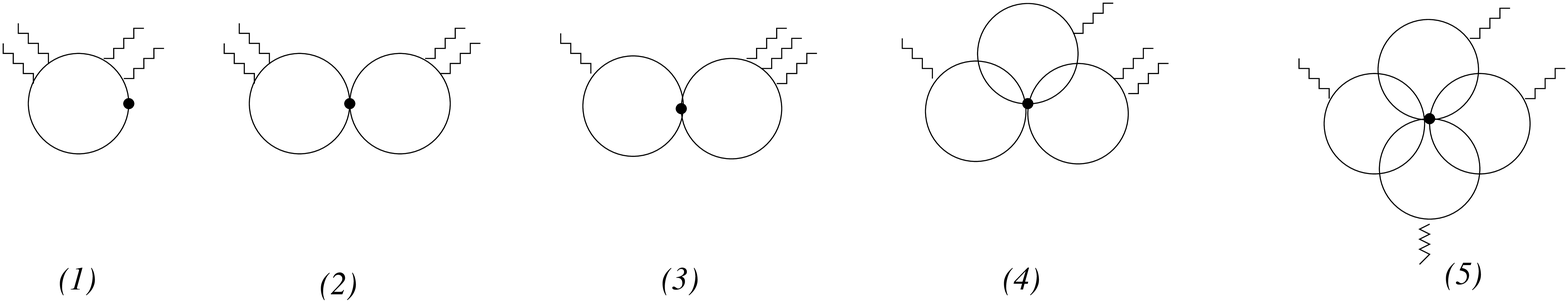} 
\includegraphics[width=13cm]{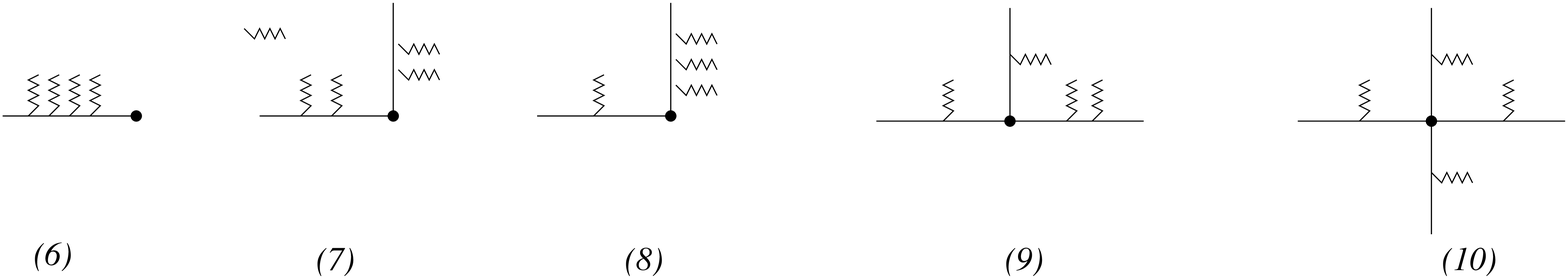} 
\includegraphics[width=13cm]{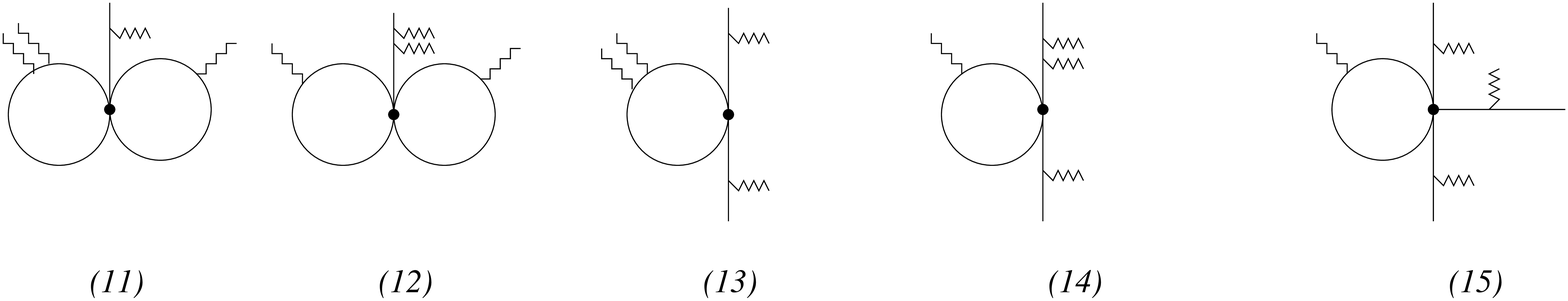} 
\includegraphics[width=13cm]{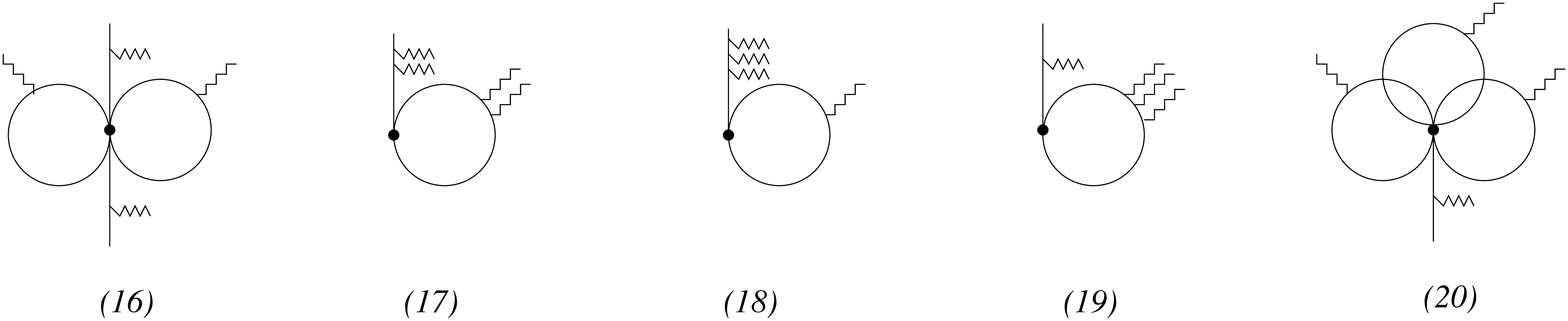} 
\end{center} 
\caption{Diagrammatic presentation of contributions into $\langle \zeta(\vec{k}_1,\vec{k}_2) \rangle$. 
The wavy lines denote restriction points $s_1$,  $s_2$, $s_3$, $s_4$. 
With $\bullet$ we denote the position of the starting ``branching'' point.} 
\label{adiag} 
\end{figure} 

The products of the components of the gyration tensor (\ref{mom}) which appear in (\ref{A}) can be calculated using the identity 
\begin{eqnarray} 
&& ({\vec{r}_i}^{\,\alpha}( s_2){-}\vec{r}_j^{\,\alpha}( s_1 ))( \vec{r}_i^{\,\beta}( s_2){-}\vec{r}_j^{\,\beta}( s_1 ) ) 
({\vec{r}_l}^{\,\alpha}( s_4){-}\vec{r}_m^{\,\alpha}( s_3 ))( \vec{r}_l^{\,\beta}( s_4){-}\vec{r}_m^{\,\beta}( s_3 ) )=\nonumber\\ 
&& =\frac{d}{d k_1^{\alpha}}\frac{d}{d k_1^{\beta}}\frac{d}{d k_2^{\alpha}}\frac{d}{d k_2^{\beta}}\, \zeta(\vec{k}_1,\vec{k}_2)\Big|_{\vec{k}_1=\vec{k}_2=0}\label{ozn} 
\end{eqnarray} 
with 
\begin{equation} 
\zeta(\vec{k}_1,\vec{k}_2)={\rm e} ^{-\vec{k}_1(\vec{r}_i(s_2)-\vec{r}_j(s_1))}{\rm e} ^{-\vec{k}_2(\vec{r}_l(s_4)-\vec{r}_m(s_3))}. 
\end{equation} 
%In what follows, we will use the notation: 
%\begin{eqnarray} 
%\zeta_{\alpha\beta|\alpha\beta}\equiv\int\!\! d s_1 \!\!\int\!\! d s_2\!\! 
% \int\!\! d s_3 \!\int\!\! d s_4 \frac{d}{d k_1^{\alpha}}\frac{d}{d k_1^{\beta}}\frac{d}{d k_2^{\alpha}}\frac{d}{d k_2^{\beta}}\, %\zeta(\vec{k}_1,\vec{k}_2)\Big|_{\vec{k}_1=\vec{k}_2=0}. \nonumber 
%\end{eqnarray} 
Again, we will use the diagrammatic presentation of contributions into $\langle \zeta(\vec{k}_1,\vec{k}_2)\rangle$ (see Fig. \ref{adiag}). 
Applying the same rules of diagram calculation, as introduced in the previous subsection and evaluating the corresponding expressions   we find   
\begin{eqnarray} 
\langle Q_{\alpha\beta}Q_{\alpha\beta}\rangle = f_1 D_{\alpha\beta|\alpha\beta}^{(1)}+\frac{f_1(f_1-1)}{2}(D_{\alpha\beta|\alpha\beta}^{(2)}+D_{\alpha\beta|\alpha\beta}^{(3)})+\nonumber\\ 
+\frac{f_1(f_1-1)(f_1-2)}{6}D_{\alpha\beta|\alpha\beta}^{(4)} 
+\frac{f_1(f_1-1)(f_1-2)(f_1-3)}{24}D_{\alpha\beta|\alpha\beta}^{(5)}+\nonumber\\+f_2 D_{\alpha\beta|\alpha\beta}^{(6)}+ 
\frac{f_2(f_2-1)}{2}(D_{\alpha\beta|\alpha\beta}^{(7)}+D_{\alpha\beta|\alpha\beta}^{(8)})+\nonumber\\ 
+\frac{f_2(f_2-1)(f_2-2)}{6}D_{\alpha\beta|\alpha\beta}^{(9)}+\frac{f_2(f_2-1)(f_2-2)(f_2-3)}{24}D_{\alpha\beta|\alpha\beta}^{(10)}+\nonumber\\ 
+\frac{f_1(f_1-1)f_2}{2}(D_{\alpha\beta|\alpha\beta}^{(11)}+D_{\alpha\beta|\alpha\beta}^{(12)})+ 
+\frac{f_1f_2(f_2-1)}{2}(D_{\alpha\beta|\alpha\beta}^{(13)}+D_{\alpha\beta|\alpha\beta}^{(14)})+\nonumber\\ 
+\frac{f_1f_2(f_2-1)(f_2-2)}{6}D_{\alpha\beta|\alpha\beta}^{(15)}+\frac{f_1(f_1-1)f_2(f_2-2)}{4}D_{\alpha\beta|\alpha\beta}^{(16)}+\nonumber\\ 
+f_1f_2(D_{\alpha\beta|\alpha\beta}^{(17)}+D_{\alpha\beta|\alpha\beta}^{(18)}+D_{\alpha\beta|\alpha\beta}^{(19)})+\frac{f_1(f_1-1)(f_1-2)f_2}{6}D_{\alpha\beta|\alpha\beta}^{(20)}. \label{qfinal} 
\end{eqnarray} 
Here, $D_{\alpha\beta|\alpha\beta}^{(n)}$ denotes contribution of $n$th diagram on Fig. \ref{adiag} (see Appendix for 
details).

\begin{figure}[b!] 
\begin{center} 
\includegraphics[width=7.0cm]{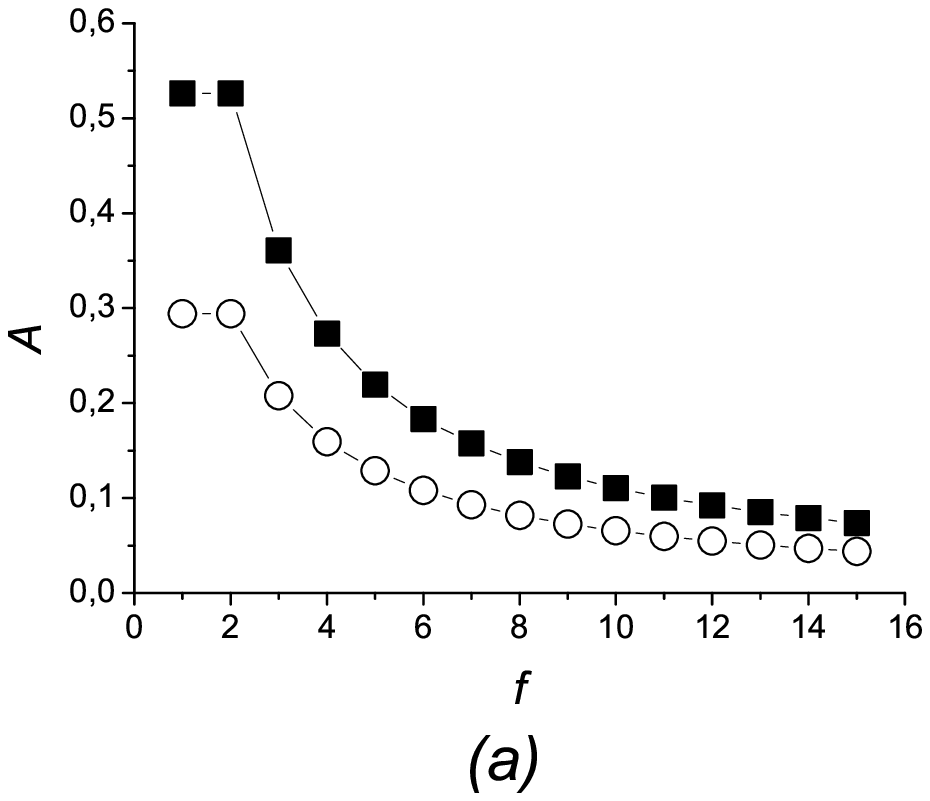} 
\includegraphics[width=7.5cm]{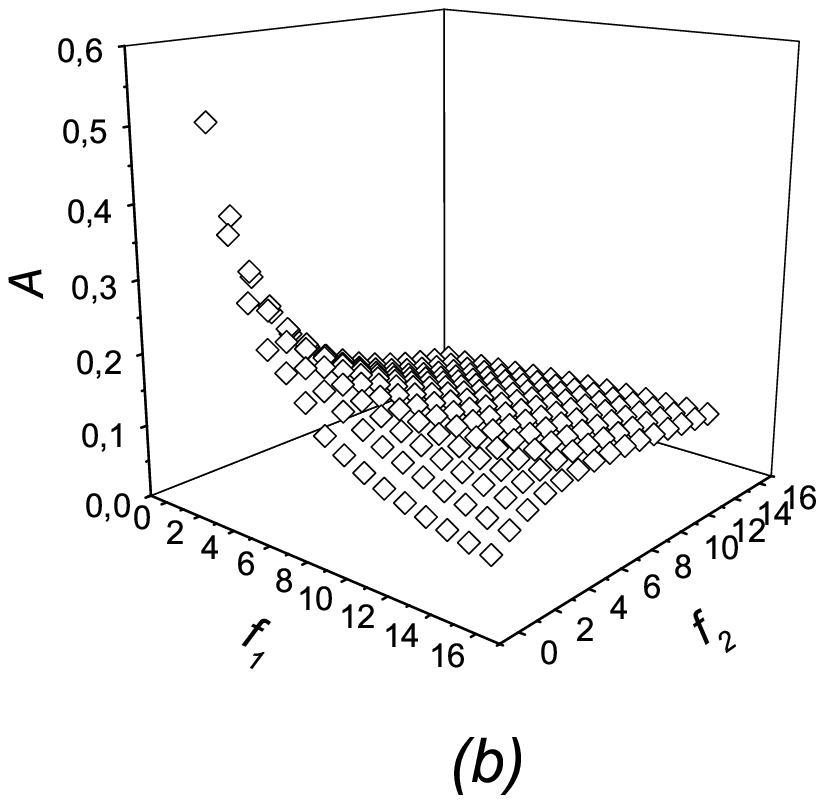}
\end{center} 
\hspace*{0.5cm} 
\caption{ \label{Adall}  (a): Asphericity 
of rosette structure (\ref{adros}) (open circles)  and star-like structure (\ref{Adstar1}) (filled squares) of the same total molecular weight 
as functions of branching parameter $f$ in space dimension $d=3$. Lines are guides to the eyes. (b):  Asphericity of the system consisting of $f_1$  polymer loops  and $f_2$ linear  branches (\ref{Adf1f2}) 
as function of $f_1$, $f_2$ in space dimension $d=3$.} 
\end{figure} 
The resulting expression for the asphericity thus reads 
\begin{eqnarray} 
&&   \hat{A}_{f_1f_2} = (2+d)\left[8f_2^2(15f_2-14)+f_1^2(8f_1-7)+\right.\nonumber\\ 
&&\left.+4f_1f_2(62f_1+28f_2-41)\right]\times \left[5d\left(4f_2^2(3f_2-2)^2+f_1^2(2f_1+1)^2+\right.\right.\nonumber\\ 
&&\left.+4f_1f_2(f_2^2(72f_2-238)-f_1^2(24f_1+113)+196)\right)+2f_1^2(8f_1-7)+\nonumber\\ 
&&\left.+16f_2^2(15f_2-14)+8f_1f_2(62f_1+34f_2-41)\right]^{-1}. \label{Adf1f2}
 \end{eqnarray} 
The case $f_2=0$ corresponds to the rosette structure (Fig. \ref{structura}a) with 
\begin{eqnarray} 
\hat{A}_{{\rm rosette}} = \frac{(2+d)(8f_1-7)}{5d(2f_1-1)^2+2(8f_1-7)}, \label{adros} 
\end{eqnarray} 
at $f_1=1$ one restores the expression of the asphericity  of an individual ring polymer (\ref{Adring}). 
{Note that for the system of two bonded rings ($f_2=2$) we restore
expression (\ref{Adring}) in accordance with the numerical results of
Ref.~\cite{Bohn10a}.}
The case $f_1=0$ corresponds to star structure (Fig. \ref{structura}c) with
\begin{equation} 
\hat{A}_{{\rm star}} = 2\frac{(2+d)(15f_2-14)}{5d(3f_2-2)^2+4(15f_2-14)}. \label{Adstar1} 
\end{equation} 
Again, at $f_2=1$, $2$ one restores an expression of asphericity  of an individual linear polymer (\ref{Adchain}).

To compare the degree of asphericity of rosette and star polymer structures of the same molecular weight (at $f_1=f_2=f$), we plot the above given 
quantities as functions of $f$ at fixed $d=3$ (see Fig. \ref{Adall}a).   
 At small $f$, the star polymers are more anisotropic and extended in space than rosette structures, 
whereas  both $\hat{A}_{{\rm star}} $ and $\hat{A}_{{\rm rosette}}$ gradually tend to zero with increasing $f$.  Really, in the asymptotic limit $f\to\infty$ both the rosette and star 
  structures can be treated as soft colloidal particles with highly symmetrical shape.   
The total asphericity of the system consisting of $f_1$ closed polymer loops and $f_2$ linear branches (Eq. (\ref{Adf1f2})) is plotted as function of $f_1$, $f_2$ at fixed $d=3$ in Fig. \ref{Adall}b. 
The value of this quantity is the result of competition of two effects: decreasing the asymmetry with increasing the number of closed loops and increasing the degree of anisotropy with increasing the number of linear branches. 

\section{Conclusions} 

In the present paper, we analysed the conformational properties of polymer systems of complex topologies
 (see Fig. \ref{structura}). 
Whereas the properties of so-called star polymers (Fig. \ref{structura}c)
have been intensively studied, much less is known about the details of  rosette (Fig. \ref{structura}a) and ring-linear (Fig. \ref{structura}b) 
structures. 
Multiple loop formation in polymer macromolecules plays an important role in  biochemical processes such as
DNA compactification  \cite{Fraser06,Simonis06,Dorier09}, which makes the rosette-like structures interesting objects to study.  Note, that another possible interpretation of structures (Fig. \ref{structura}a) and  (Fig. \ref{structura}b) is the following: they can be treated as projections of long flexible polymer in the bulk 
onto the 2-dimensional plane \cite{Metzler02}. 

Restricting ourselves to the idealised Gaussian case, when any interactions between monomers are neglected, we develop
the continuous chain representation of the complex polymer model, considering each of the individual branches or petals 
as a path of length $S_i$, parameterised by $\vec{r}_i(s)$, where $s$ is varying from $0$ to $S_i$ ($i=1,2,\ldots,f_1+f_2$). 
The size and shape characteristics of a typical polymer conformation have been studied on the basis 
of the gyration tensor $\bf{Q}$ with the 
 the  components given by 
 (\ref{mom}). Working within the framework of path integration method and making use of appropriate diagram technique, 
 we obtained the expressions for the gyration radius $\langle R^2_{g\,f_1f_2} \rangle$ and asphericity ${\hat {A}}$, measuring 
the extent of anisotropy 
of a typical conformation of  complex polymer structures, as functions of parameters $f_1$, $f_2$. 
{In particular, our analytical results quantitatively confirm the
compactification (decrease of the effective size) of multiple loop polymer
structures as compared with structures containing linear segments
[Eq.~(\ref{ratioRS})]. A decrease of the anisotropy of rosette polymers as
compared to star-like structures of the same total molecular weight is revealed,
as well [Eqs.~(\ref{adros}), (\ref{Adstar1})].}

\section*{Acknowledgements} 

This work was supported  by the 
FP7 EU IRSES projects N269139  ``Dynamics and Cooperative Phenomena in Complex 
Physical and Biological Media'' and N295302 ``Statistical Physics in Diverse Realizations''. 
 Financial support from the Academy 
 of Finland within the FiDiPro scheme is acknowledged. 

\section*{Appendix} 
\begin{figure}[t!] 
\begin{center} 
\includegraphics[width=4cm]{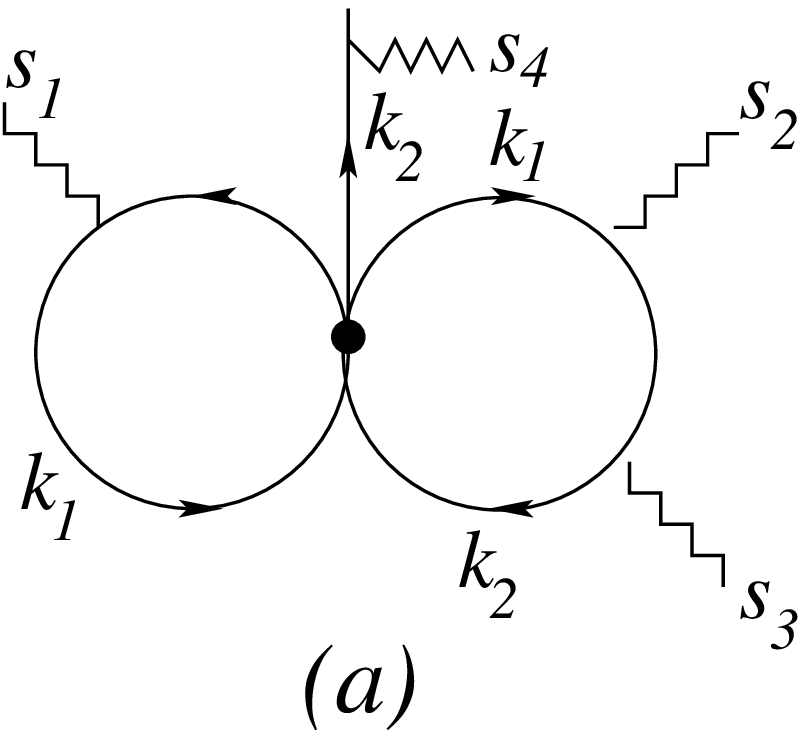} 
\hspace*{0.2cm}\includegraphics[width=4.4cm]{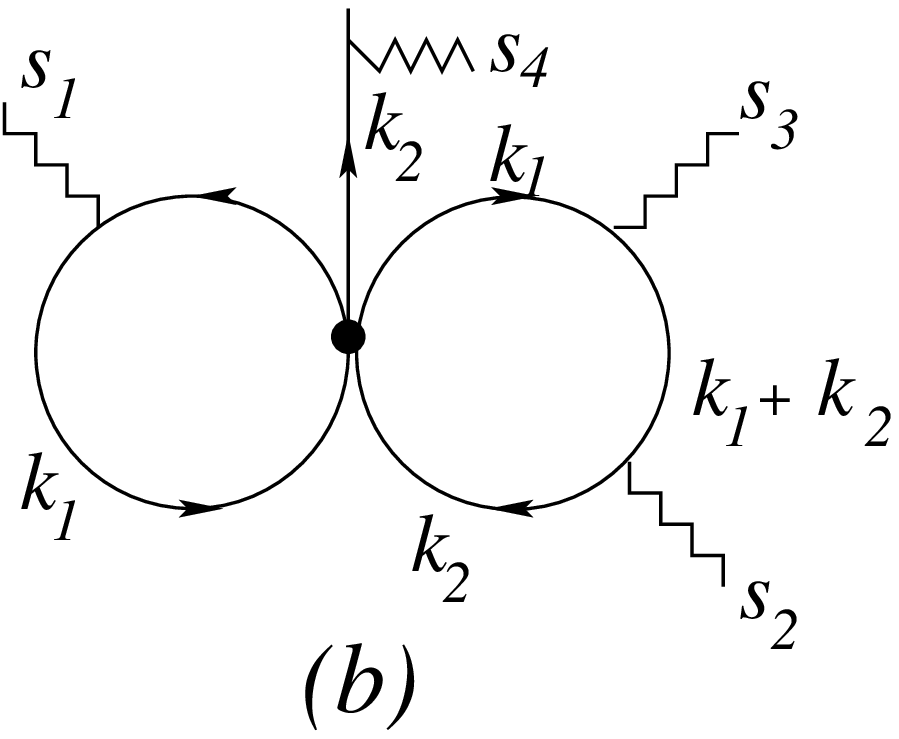} 
\hspace*{0.2cm}\includegraphics[width=4cm]{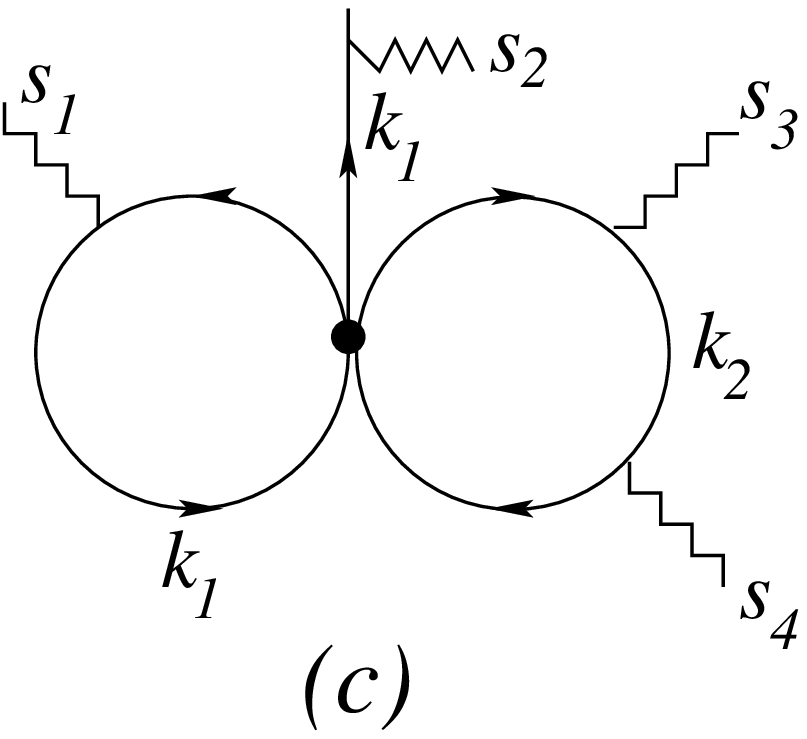} 
\includegraphics[width=4cm]{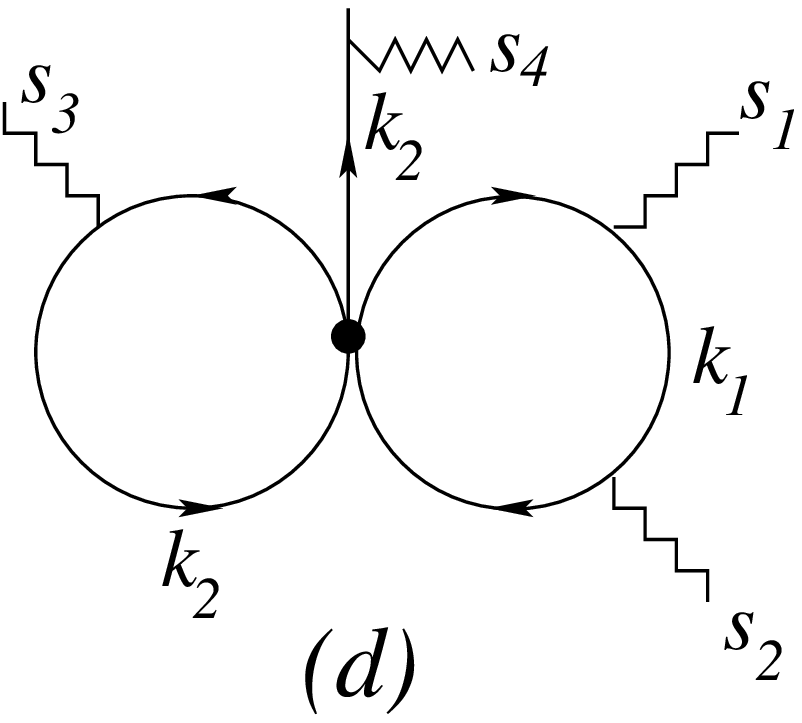} 
\hspace*{0.2cm}\includegraphics[width=4.4cm]{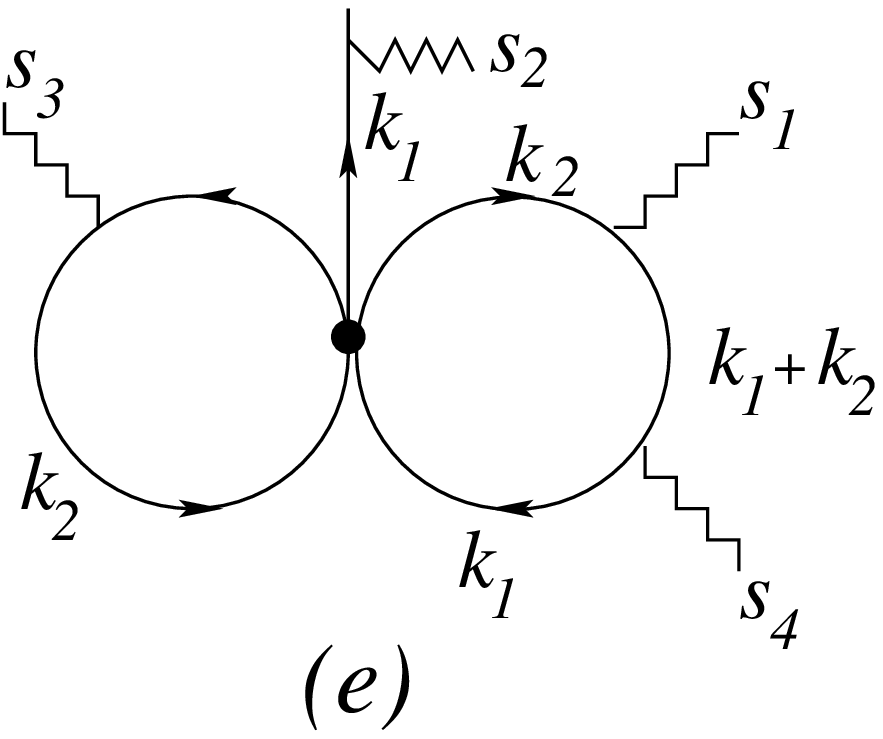} 
\hspace*{0.2cm}\includegraphics[width=4cm]{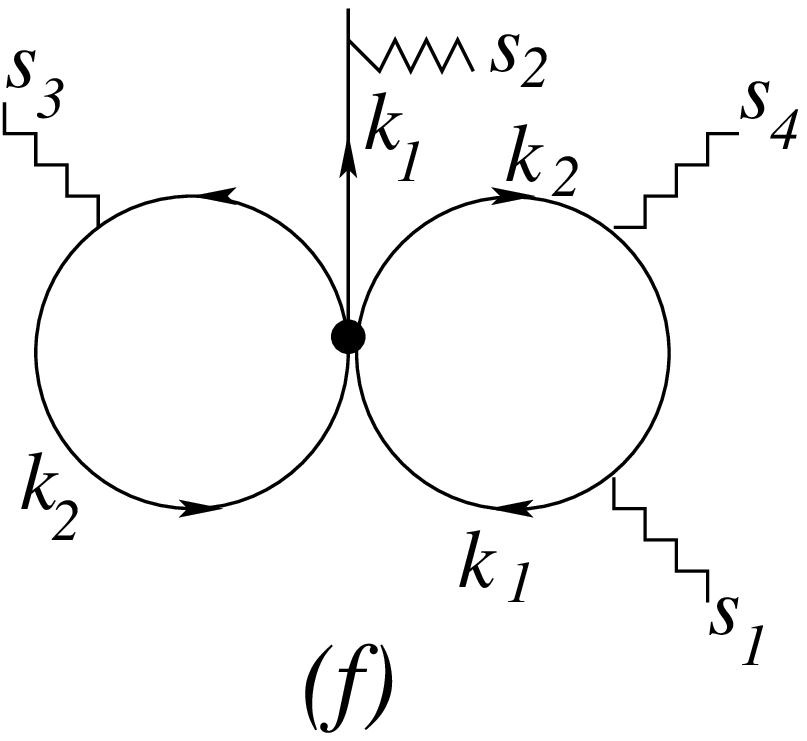} 
\end{center} 
\caption{Diagram (11) on Fig. \ref{adiag} with various permutations of restriction points $s_1$,  $s_2$, $s_3$, $s_4$.} 
\label{11} 
\end{figure}

Here, we evaluate the analytical expressions, corresponding to diagram (11) on Fig. \ref{adiag}. Note, that in the diagram calculations one 
should take into account the possible permutations of positions of restriction points $s_1$,  $s_2$, $s_3$, $s_4$ (see Fig. \ref{11}). 

  The resulting expressions read: 
 \begin{eqnarray} 
 \zeta^{(11)a}(\vec{k}_1,\vec{k}_2)={\rm e}^{-\frac{k_1^2}{2}(S-s_1+s_2)}{\rm e}^{-\frac{k_2^2}{2}(S-s_3+s_4)}\! \!\int\!\!{\rm d}q_1\!\!\int\!\!{\rm d}q_2\, 
  {\rm e}^{-\frac{q_1^2}{2}S}{\rm e}^{-\frac{q_2^2}{2}S}\times \nonumber\\ 
  \times{\rm e}^{-{q_1k_1}(S-s_1)}
  {\rm e}^{-{q_2k_1}s_2} {\rm e}^{-{q_2k_2}(S-s_3)} = \nonumber\\ 
  = {\rm e}^{-\frac{k_1^2}{2S}(S(s_1+s_2)-s_1^2-s_2^2)-\frac{k_2^2}{2S}(S(s_3+s_4)-s_3^2) +\frac{k_1k_2}{S}s_2(S-s_3)},\nonumber\\ 
   \zeta^{(11)b}={\rm e}^{-\frac{k_1^2}{2}(S-s_1+s_2)}{\rm e}^{-\frac{k_2^2}{2}(S-s_3+s_4)} {\rm e}^{-k_1k_2(s_2-s_3)}  \! \!\int\!\!{\rm d}q_1\!\!\int\!\!{\rm d}q_2\, 
  {\rm e}^{-\frac{q_1^2}{2}S}{\rm e}^{-\frac{q_2^2}{2}S}\times\nonumber \\
  \times{\rm e}^{-{q_1k_1}(S-s_1)} 
  {\rm e}^{-{q_2k_1}s_2} {\rm e}^{-{q_2k_2}(S-s_3)} = \nonumber \\ = {\rm e}^{-\frac{k_1^2}{2S}(S(s_1+s_2)-s_1^2-s_2^2)-\frac{k_2^2}{2S}(S(s_3+s_4)-s_3^2) +\frac{k_1k_2}{S}(s_2(S-s_3)-S(s_2-s_3))},\nonumber\\ 
   \zeta^{(11)c}={\rm e}^{-\frac{k_1^2}{2}(S-s_1+s_2)}{\rm e}^{-\frac{k_2^2}{2}(s_4-s_3)} \! \!\int\!\!{\rm d}q_1\!\!\int\!\!{\rm d}q_2\, 
  {\rm e}^{-\frac{q_1^2}{2}S}{\rm e}^{-\frac{q_2^2}{2}S}{\rm e}^{-{q_1k_1}(S-s_1)}\times \nonumber\\ 
  \times {\rm e}^{-{q_2k_2}(s_4-s_3)} =  {\rm e}^{-\frac{k_1^2}{2}(s_1+s_2)- 
  \frac{k_2^2}{2S}(s_4-s_3)(S-s_4+s_3)},\nonumber\\ 
 \zeta^{(11)d}={\rm e}^{-\frac{k_1^2}{2}(s_2-s_1)}{\rm e}^{-\frac{k_2^2}{2}(S-s_3+s_4)} \! \!\int\!\!{\rm d}q_1\!\!\int\!\!{\rm d}q_2\, 
  {\rm e}^{-\frac{q_1^2}{2}S}{\rm e}^{-\frac{q_2^2}{2}S}{\rm e}^{-{q_1k_2}(S-s_3)}\times \nonumber\\ 
  \times {\rm e}^{-{q_2k_1}(s_2-s_1)} 
  =  {\rm e}^{-\frac{k_1^2}{2S}(s_2-s_1)(S-s_2+s_1)-\frac{k_2^2}{2}(s_3+s_4)},\nonumber\\ 
   \zeta^{(11)e}={\rm e}^{-\frac{k_1^2}{2}(S-s_1+s_2)}{\rm e}^{-\frac{k_2^2}{2}(S-s_3+s_4)} {\rm e}^{-k_1k_2(s_4-s_1)} \! \!\int\!\!{\rm d}q_1\!\!\int\!\!{\rm d}q_2\, 
  {\rm e}^{-\frac{q_1^2}{2}S}{\rm e}^{-\frac{q_2^2}{2}S}\times \nonumber\\ \times{\rm e}^{-{q_1k_2}(S-s_3)}
  {\rm e}^{-{q_2k_1}(S-s_1)} {\rm e}^{-{q_2k_2}s_4} = \nonumber\\ = {\rm e}^{-\frac{k_1^2}{2S}(S(s_1+s_2)-s_1^2)-\frac{k_2^2}{2S}(S(s_3+s_4)-s_3^2-s_4^2) +\frac{k_1k_2}{S}(s_4(S-s_1)-S(s_4-s_1))},\nonumber\\ 
   \zeta^{(11)f}={\rm e}^{-\frac{k_1^2}{2}(S-s_1+s_2)}{\rm e}^{-\frac{k_2^2}{2}(S-s_3+s_4)} \! \!\int\!\!{\rm d}q_1\!\!\int\!\!{\rm d}q_2\, 
  {\rm e}^{-\frac{q_1^2}{2}S}{\rm e}^{-\frac{q_2^2}{2}S}{\rm e}^{-{q_1k_2}(S-s_3)}\times \nonumber\\\times {\rm e}^{-{q_2k_2}s_4}
   {\rm e}^{-{q_2k_1}(S-s_1)} = \nonumber \\ {\rm e}^{-\frac{k_1^2}{2S}(S(s_1+s_2)-s_1^2)-\frac{k_2^2}{2S}(S(s_3+s_4)-s_4^2) 
  +\frac{k_1k_2}{S}s_4(S-s_1)}. \nonumber 
 \end{eqnarray} 
 Taking derivatives over the components of $k_1$, $k_2$ in expressions above according to (\ref{ozn}) and integrating over $s_1,\ldots,s_4$, 
one finally obtains the contributions $D_{\alpha\beta|\alpha\beta}^{(11)}$ in (\ref{qfinal}):   
 \begin{equation} 
 D_{\alpha\alpha|\alpha\alpha}^{(11)}=\frac{59}{45}S^6,\,\,\,\,\, D_{\alpha\alpha|\beta\beta}^{(11)}=\frac{11}{9}S^6,\,\,\,\,\,D_{\alpha\beta|\alpha\beta}^{(11)}=\frac{2}{45}S^6. 
 \end{equation}

\end{document}